\documentclass[prc,twocolumn,superscriptaddress,showpacs,preprintnumbers,amsmath,amssymb]{revtex4}
\usepackage[dvips]{graphicx}
\usepackage{dcolumn}
\def\m@thcombine#1#2{%
  \setbox0=\hbox{$#1$}
  \setbox1=\hbox{$#2$}
  \ifdim\wd0>\wd1
    \setbox0=\hbox to\wd1{\hss\box0\hss}
  \else
    \setbox1=\hbox to\wd0{\hss\box1\hss}
  \fi
  \mathop{\vcenter{
    \offinterlineskip\box0\box1}}}
\def\lesim{\m@thcombine<\sim}
\def\gesim{\m@thcombine>\sim}

\begin{document}
\title{Emergence of pygmy dipole resonances: Magic numbers and neutron skins}
\author{Tsunenori Inakura}
\affiliation{
RIKEN Nishina Center, Wako, 351-0198, Japan}
\author{Takashi Nakatsukasa}
\affiliation{
RIKEN Nishina Center, Wako, 351-0198, Japan}
\affiliation{Center for Computational Sciences, University of Tsukuba, Tsukuba 305-8571, Japan}
\author{Kazuhiro Yabana}
\affiliation{Center for Computational Sciences, University of Tsukuba, Tsukuba 305-8571, Japan}
\affiliation{
RIKEN Nishina Center, Wako, 351-0198, Japan}

\begin{abstract}
The pygmy dipole resonances (PDR) for even-even nuclei in $8\leq Z \leq 40$
are studied performing a systematic calculation of
the random-phase approximation with the Skyrme functional of SkM*.
The calculation is fully self-consistent and does not assume any symmetry
in the nuclear shape of the ground state.
In every isotopic chain,
the PDR emerges by showing
a peak of the $E1$ strength at energies less than 10 MeV.
The $E1$ strength of the PDR strongly depends on the position of
the Fermi level and shows a clear correlation with
the occupation of the orbits with the orbital angular momenta less than
$3\hbar$ ($\ell\leq 2$).
We also found a strong correlation between the isotopic dependence of
the neutron skin thickness and the pygmy dipole strength.
The fraction of the energy weighted strength exhausted
by the PDR and the neutron skin thickness
show a linear correlation with the universal rate of
about 0.2 fm$^{-1}$.
\end{abstract}

\maketitle

Exotic nuclei show interesting new features in their response properties.
The existence of loosely bound nucleons in the neutron skin and halo
will significantly modify the strength distributions.
Among them, the low-energy electric
dipole ($E1$) excitation, which
emerges at the energy well below the giant dipole resonance (GDR),
has been most extensively studied so far.
These low-energy peaks in the $E1$ strength distribution
are often called the pygmy dipole resonance (PDR).
The PDR properties have a strong impact on astrophysical phenomena.
The low-energy $E1$ strengths in neutron-rich nuclei strongly enhance
the radiative capture cross sections for low-energy
neutrons, which directly affects the abundance 
distribution in the rapid neutron capture process \cite{Goriely}.
In contrast to the GDR, the PDR is sensitive to 
nuclear properties at nuclear surface and at low density.
Thus, its property may provide us with useful constraints on
the energy density functional,
to identify the equation of state (EOS) of
the nuclear and neutron matters.

Another intriguing subject of interest is
the relation between the PDR and the neutron skin thickness.
The neutron skin thickness is known to be well correlated with
the EOS of the neutron matter \cite{Brown}.
Thus, if the PDR is correlated with the neutron skin thickness,
the investigation of the PDR may reveal the properties of
the neutron stars, such as the proton ratio and radius.
Piekarewicz suggested that the PDR strength
is strongly correlated with the neutron skin thickness
\cite{Piekarewicz}.
This means that the measurement of the PDR can provide information on the skin 
thickness.
In contrast, a recent correlation analysis, which investigates
the parameter dependence of various properties in $^{208}$Pb,
suggests a very weak
correlation between the skin thickness and the PDR \cite{Reinhard}.
These conclusions were, however, obtained from the calculations for
specific nuclei, such as
spherical Sn isotopes and $^{208}$Pb.
It is thus desirable to perform a systematic and fully self-consistent
calculation for the PDR for spherical and deformed nuclei,
to reveal the nature of the PDR and its relation to the neutron skin.

The PDR were 
observed experimentally in neutron-rich nuclei such as O isotope 
\cite{Leistenschneider,Tryggestad}, $^{26}$Ne \cite{Gibelin}, $^{68}$Ni 
\cite{Wieland} and Sn isotopes \cite{Adrich}, but also in stable Ca 
\cite{Hartmann}, Sn \cite{Govaert,Klimkiewicz} and Pb isotopes 
\cite{Ryezayeva,Enders}, and $N=82$ isotones
\cite{Herzberg,Zilges,Volz,Savran,Tonchev} as well.
Although the nature of the PDR, including its relation to the neutron
skin and halo, is still elusive,
these experiments have identified the following
properties of the PDR:
The concentrated $E1$ strengths 
have been observed around the threshold energy for the neutron separation. 
The energy-weighted strength for the PDR is less than 
$1 \%$ of the Thomas-Reiche-Kuhn (TRK) sum-rule value in the stable nuclei and 
less than $5 \%$ in the case of neutron-rich nuclei.

Many microscopic studies based on the random 
phase approximation (RPA) with the Skyrme interaction 
\cite{Colo,Sarchi,Terasaki,Matsuo,Yoshida,Ebata}, the Gogny interaction 
\cite{Peru,Martini}, and those based on the relativistic mean-field approach 
\cite{Vretenar,Paar,Piekarewicz,Liang,Pena} 
were carried out to investigate the PDR in the past decade.
These microscopic calculations succeeded, to some extent,
to reproduce dipole strength at low excitation.
However, these 
microscopic calculations were applied to limited number of nuclei so far,
and even basic questions, such as the condition on which the PDR appears in
a nucleus, are not fully understood yet.

In the present paper,
we report our systematic calculation of the $E1$ response for a wide 
mass region with $8\leq Z \leq 40$ (up to $A \sim 110$)
in the self-consistent RPA with the Skyrme 
interaction, focusing on the emergence and the properties of the PDR.
In the present calculation,
we do not take into account the pairing correlation because we expect 
that its effect is not significant for the $E1$ response in light- and 
medium-mass nuclei \cite{Ebata,Piekarewicz}.
We adopt the representation of
the three-dimensional (3D) Cartesian grids.
The real-space
representation has an advantage over other basis representations,
such as the harmonic oscillator basis,
on the treatment of the continuum, since it is well controlled by
the box size and can be directly compared with the continuum RPA
calculations.
We adopt the 3D adaptive coordinate grids \cite{Nakatsukasa05}
within a sphere of a radius 
$R_\mathrm{box}=15$ fm.
The Skyrme functional of the SkM* parameter set
\cite{Bartel} is used. 
All the single-particle wave functions and potentials except for the Coulomb 
potential are assumed to vanish outside the sphere.
The differentiation is approximated by a finite difference with the nine-point 
formula.

The first step of the investigation is to construct the Hartree-Fock
ground state, using the imaginary-time method.
Then, we solve the linear-response equation for the $E1$
external field directly in the
coordinate representation, avoiding explicit construction
of unoccupied orbitals. Therefore, our calculation is free
from a truncation for the unoccupied orbitals.
Since the explicit construction of the RPA matrix in the 3D mesh 
representation is difficult, we use a new 
methodology, the finite amplitude method (FAM) \cite{Nakatsukasa07},
in evaluation of the residual field $\delta h$.
The FAM allows
us to construct $\delta h$ using the calculation
of the Hamiltonian $h$ only.
The residual field $\delta h$ in our calculations contains all terms of the 
Skyrme interaction, i.e., the residual spin-orbit interaction, the time-odd 
components, the residual Coulomb interaction, and so on.
The linear-response equation is 
solved at given complex energies $\omega=E + i\gamma /2 $
using iterative solvers,
such as the 
generalized conjugate residual (GCR) method.
The calculations are 
performed at energies in spacing of $\Delta E=0.3$ MeV
with 
a fixed imaginary part 0.5 MeV, corresponding to the smearing width
$\gamma=1.0$ MeV. Using the 
obtained RPA amplitudes, we compute the $E1$ strength and the photoabsorption 
cross section.
Details of the calculation can be found in Ref. \cite{Inakura09}. 

The $E1$ strength of even-even nuclei are calculated up to zirconium isotopes
from the proton to neutron drip lines, except for nuclei
with the neutron separation energy less than 2 MeV.
This excludes the neutron drip-line nuclei in which
extended halos develop.
In addition, the neutron-deficient 
nuclei from the proton drip-line to $N=50$ are also calculated
for $40<Z\leq 50$.
The total number of the calculated 
nuclides are 322: 40 spherical nuclei, 171 prolate nuclei, 56 oblate nuclei, 
and 55 triaxial nuclei. 

\begin{figure}[tb]
\begin{center}
\includegraphics[width=0.499\textwidth,keepaspectratio]{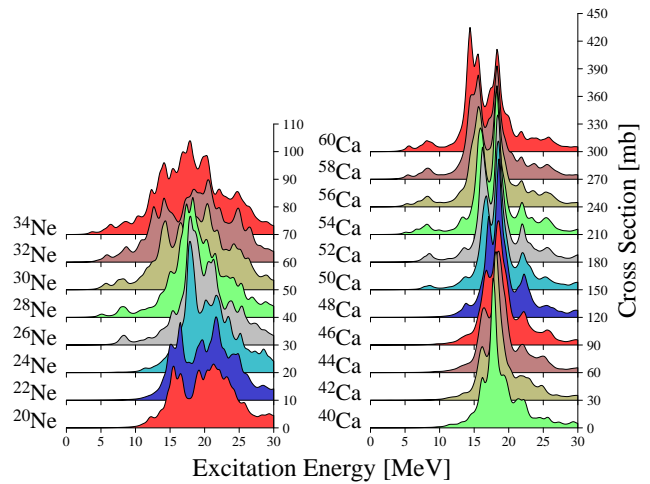}
\caption{{\small (Color online) Calculated photoabsorption cross sections in Ne and Ca isotopes.}}
\label{pygmy.example}
\end{center}
\end{figure}
\begin{figure}[tb]
\begin{center}
\includegraphics[width=0.499\textwidth,keepaspectratio]{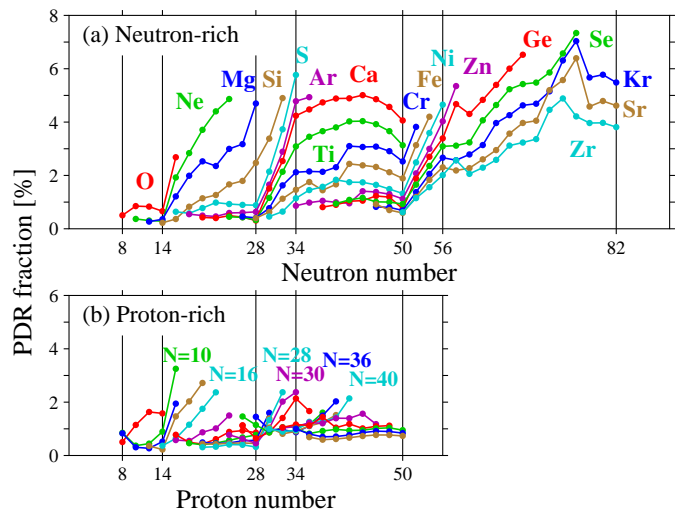}
\caption{{\small (Color online) Fractions of photoabsorption cross sections 
$m_1(\mbox{PDR})/m_1$ for (a) stable and neutron-rich nuclei and for (b)
stable and proton-rich nuclei,
as functions of neutron number and proton number, respectively.}}
\label{PDRfraction}
\end{center}
\end{figure}

The calculation shows that the PDR peaks appear in every isotopic chain,
as demonstrated in Fig. 
\ref{pygmy.example} for Ne and Ca isotopes.
Increasing neutron number with a fixed proton number,
we see that the emergence of the PDR
suddenly takes place at $N=16$ for Ne and at $N=30$ for Ca isotopes.
In Ca isotopes, the PDR is rather distinctive and separated from the
GDR peak.
However, in the deformed neutron-rich Ne isotopes, it is not well
separate from the low-energy tail of the GDR.
Although it is not trivial how to define the PDR, the separate low-energy peaks
mostly appear at energies below 10 MeV.
Thus, in this work, the pygmy dipole strength is defined 
by the $E1$ strength at energies below 10 MeV.

We calculate fractions of the photoabsorption cross section $\sigma(E)$
integrated up to $E=10$ MeV to the integrated total cross section.
This is equal to the ratio $m_1(\mbox{PDR})/m_1$
where $m_1(\mbox{PDR})$ is the energy-weighted sum up to 10 MeV and
$m_1$ is the energy-weighted-sum-rule value that is larger than
the TRK value by $30\sim 40$ \%.
The upper panel of Fig. \ref{PDRfraction} shows the PDR
fractions of stable and neutron-rich nuclei as functions of neutron number. 
Here, we show isotopes with $8\leq Z \leq 40$.
In each isotopic chain,
nuclei around stable region have small values of the fraction 
less than $\sim$1\%, consistent with the experimental data
\cite{Leistenschneider,Tryggestad,Hartmann}.
In these light nuclei, a prominent PDR 
does not appear in stable nuclei.
Being away from the stable region,
the fractions suddenly increase at specific neutron numbers:
$N>14$, $N>28$, and $N>50$.
These ``magic'' numbers for the emergence of the PDR 
indicate the presence of a strong shell effect of neutrons.
The first clear indication of the kink structure appears
at $N=14\rightarrow 16$ for O, Ne and Mg isotopes.
For these isotopes,
the strength of the PDR has a strong correlation with
the number of neutrons occupying
the orbits with low orbital angular momenta (low-$\ell$),
$s_{1/2}$ and $d_{3/2}$ ($N=15\sim 20$).
In case that the low-$\ell$ neutron orbits are weakly bound,
they strongly expand in space due to the low centrifugal barrier.
The kink is weakened by increasing the proton number and almost disappears
for Si isotopes ($Z=14$) in which $N=16$ corresponds to the stable
nucleus $^{30}$Si.
This should be due to the fact that these low-$\ell$ neutron orbits become
more bound for nuclei with larger $Z$.
The neutrons start filling the $f_{7/2}$ orbits at $N>20$, then,
the growing rate of the PDR strength is reduced, which is most prominent
for Mg isotopes.

The even more prominent kinks can be identified at the magic numbers of
$N=28$ and $N=50$.
The PDR fractions suddenly increase at $N=28\rightarrow 30$ and 
continue to increase till $N=34$ where $2p$ shell are filled.
The increasing rate of the PDR fractions depends on the proton number.
Namely, it is the largest for small-$Z$ isotopes, such as Si and Ar isotopes,
while increasing $Z$ makes the rate smaller.
Beyond $N=34$,
the PDR fractions are roughly constant for $34<N\leq 50$, in which
the neutrons are filling high-$\ell$ orbits of $f_{5/2}$ and $g_{9/2}$.
They again show a sudden increase at $N=50\rightarrow 52$, then,
continue to linearly grow up until $2d_{5/2}$ orbits are filled at $N=56$.
Beyond that, it is difficult to see the definite trend,
since the ordering of the
orbits and the ground-state deformation change
from nucleus to nucleus, depending on $Z$ and $N$.
However, the careful investigation suggests that the occupation of
$s_{1/2}$ and $d_{3/2}$ increases the PDR fraction,
while that of $h_{11/2}$ reduces it.
Thus, we may conclude that the spatially extended nature of the low-$\ell$
neutron orbits near the Fermi level plays a primary role for the
the emergence and growth of the PDR.
We have also observed that the deformation tends to increase the PDR
strength, especially in the region $N>56$.
This may be due to two effects;
the mixture of the low-$\ell$ components in the orbits near the Fermi
level and softening of the giant dipole resonance for the
elongated direction.

In the proton-rich side, we can see the similar behaviors 
for the isotones with $8\leq N\leq 50$
in the lower panel of Fig. \ref{PDRfraction}.
There are kinks in the PDR fractions at $Z=14$ and 28.
However, the increase as a function of the proton number at $28<Z\leq 34$ is 
less prominent than that as a function of the neutron number
at $28<N\leq 34$.
This is because the Coulomb barrier prevents the 
low-$\ell$ proton orbits from spatial extension, especially in 
high-$Z$ nuclei.

Assuming the Steinwedel-Jensen model with the core and valence
neutrons (protons), 
the PDR fraction is proportional to $N_v/N$ ($Z_v/Z$) where $N_v$ ($Z_v$)
is the number of valence neutrons (protons) \cite{Suzuki}.
This is consistent with our result shown in Fig. \ref{PDRfraction} for
valence nucleons defined as those in the low-$\ell$ orbits.
However, the model predicts the PDR peak energy as
$E_{\rm PDR}/E_{\rm GDR}=\sqrt{N_v/N}$,
which disagree with our microscopic result.
Figure \ref{pygmy.example} suggests that the PDR peak energy is rather
insensitive to the number of nucleons.

\begin{figure}[tb]
\begin{center}
\includegraphics[width=0.450\textwidth,keepaspectratio]{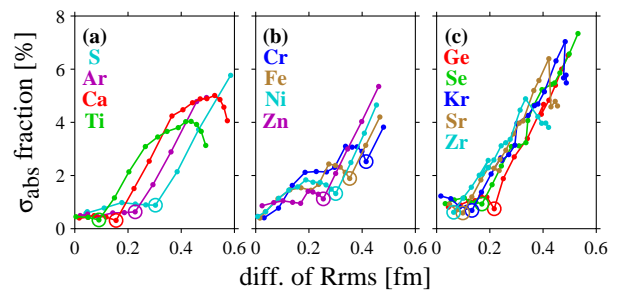}
\caption{{\small (Color online) Correlations between fractions of photoabsorption cross section and neutron skin thickness.
The kink positions at $N=28$ and 50 are denoted by circles.
See text for details.
}}
\label{PDRcorr}
\end{center}
\end{figure}

Next, let us examine the correlation between the PDR and the neutron
skin thickness. 
The skin thickness is defined by the difference in radius between
neutrons and protons: $\Delta R_{\rm rms}\equiv\sqrt{\langle r_n^2 \rangle} -
\sqrt{\langle r_p^2 \rangle}$.
We find that $\Delta R_{\rm rms}$ increases
linearly with respect to the number of valence neutrons
in the regions of $N=14-20$, $N=28-34$, and $N=50-56$,
where the PDR fraction also shows significant linear increase
as a function of the neutron number (Fig.~\ref{PDRfraction}).
Plotting the PDR fraction as a function of $\Delta R_{\rm rms}$,
we then observe a linear correlation between them.
This is illustrated in Fig. \ref{PDRcorr} for isotopes with $Z=16\sim 40$
which show the kinks at $N=28$ and 50.
The PDR fraction in each isotopic chain shows a linear correlation with
the skin thickness in the regions of the neutron number
$N=28-34$ and $N \ge 50$.
The positions of the kinks are located at different values of
$\Delta R_{\rm rms}$ for different isotopes.
However, the slope is universal for all the isotopes;
$0.18\sim 0.20$ fm$^{-1}$.
It is remarkable to see that, even though the PDR fractions show
no linear dependence on the neutron number at $N>56$ in
Fig. \ref{PDRfraction}, they still keep the linear correlation with
the skin thickness, until they start to decrease due to
the dominant role of the $h_{11/2}$ orbits
at $N\gtrsim 76$.
Despite the fact that the deformation and shell ordering are different and
vary from nucleus to nucleus,
the universal linear correlation remains valid for $50\leq N \lesssim 76$.
It should be noted that the linear correlation can be observed
 only for each isotopic chain.
Deleting the lines connecting isotopic chains in Fig. \ref{PDRcorr},
we only see scattered points showing a weak correlation.

A detailed comparison with the available experimental data is beyond
the scope of the present paper, since we use a smearing width of $\gamma=1$
MeV which prevents us from quantifying small low-energy $E1$ strengths
which carry less than 1 \% of the TRK sum value.
So far, the observed PDR strength in nuclei with $Z\leq 40$
are very small (typically less than 1 \% of the TRK value),
which is qualitatively consistent with our calculation because these nuclei
are in the region before the kinks ($N\leq 14$, 28, 50).
The only exception is $^{26}$Ne \cite{Gibelin}.
The experimental data for $^{26}$Ne indicates
the PDR peak at $E_{\rm PDR}\approx 9$ MeV carrying
about 5 \% of the TRK value.
These values correspond to our calculated values
of $E_{\rm PDR}\approx 8$ MeV with 2.6 \%.
The low-energy $E1$ strength in $^{68}$Ni is observed around $E=11$ MeV
carrying about 5 \% of the TRK value \cite{Wieland}.
Our calculation predict a PDR peak around $E=10$ MeV and
$m_1(\mbox{PDR})/m_1$ indicates 4.6 \% of the TRK value
where $m_1(\mbox{PDR})$ is defined by the strength summed up to $E=12$ MeV.

Finally, we comment on dependence of the results on the box size $R_{\rm box}$
of the calculation.
We have selected several nuclei to check how the result depends on
 $R_{\rm box}$.
The discretized continuum states, which depend on the choice of $R_{\rm box}$,
may affect the precise
peak position and shape of the PDR, however,
the PDR strength is very little affected.
For instance,
the enlargement of the $R_{\rm box}$ to 25 fm shifts the PDR
fractions, 1.89\% $\to$ 1.94\% for $^{26}$Ne and 4.79\% $\to$ 
4.65\% for $^{60}$Ca.
Thus, the features of the PDR discussed above are
valid in the present calculation.

In summary,
we carried out the systematic calculation of the low-lying $E1$ strengths 
in even-even nuclei up to zirconium isotopes,
including both spherical and deformed nuclei for the first time,
by means of the fully self-consistent RPA 
with the Skyrme interaction utilizing the FAM.
We found the PDR in every isotopic chain.
The PDR strength in neutron-rich nuclei shows strong enhancement
in $N>14$ for nuclei with $7<Z<13$, in $N>28$ for $13<Z<23$, and
in $N>50$ for $23<Z\leq 40$.
This suggests that the occupation of the low-$\ell$,
such as $s$, $p$ and $d$ orbits, play 
a key role for the emergence of the PDR in neutron-rich nuclei.
In the proton-rich side, the PDR strengths are hindered
because of the Coulomb barrier to prevent those orbits from the spatial
extension.
The PDR distributions show a remarkable linear correlation with the 
neutron skin thickness in the regions of $28\leq N\leq 34$ and
$50\leq N\lesssim 76$,
with a universal slope of $0.18\sim 0.20$ /fm.
The present result is consistent with the linear correlation 
reported in Ref. \cite{Piekarewicz} for neutron-deficient Sn isotopes
in the region of $56<N<70$,
where the neutrons are filling the $s$ and $d$ orbits.
Since heavy isotopes possess the neutron skin even in the neutron-deficient
side, it is of great interest to perform a systematic study on the PDR
in heavy systems.

This work is supported by the PACS-CS project of 09-a-25, 09-b-9, 10a-22, of the Center 
for Computational Sciences, University of Tsukuba and by the Large Scale 
Simulation Program No. 08-14, 09-16, and 09/10-10 of High Energy Accelerator 
Research Organization (KEK) and by Grant-in-Aid for Scientific Research on 
Innovative Areas (No. 20105003) and by the Grant-in-Aid for Scientific 
Research(B) (No. 21340073).
The numerical calculations were performed on 
PACS-CS supercomputers in University of Tsukuba, on Hitachi SR11000 at KEK, 
and on the RIKEN Integrated Cluster of Clusters (RICC).
We also thank the International 
Research Network for ``Exotic Femto Systems'' (EFES) of the Core-to-Core 
Programs of Japan Society for the Promotion of Science.

\end{document}